\documentclass[a4paper]{spie}

\usepackage[]{graphicx}

\title{Quantum points/patterns, Part 1.\\
 From geometrical points to quantum points in a sheaf framework}

\author{Antonina N.~ Fedorova\supit{a} and  Michael G.~Zeitlin\supit{a}
\skiplinehalf
\supit{a}IPME RAS, V.O. Bolshoj pr., 61, 199178, St.~Petersburg, Russia
}     
\authorinfo{anton@math.ipme.ru, zeitlin@math.ipme.ru,
http://www.ipme.ru/zeitlin.html,                                                          
http://mp.ipme.ru/zeitlin.html}

\begin{document} 

\begin{center}
\begin{tabular}{p{160mm}}

\begin{center}
{\bf\Large
Quantum points/patterns, Part 1.} \\
\vspace{5mm}

{\bf\Large  From geometrical points to quantum}\\
\vspace{5mm}

{\bf\Large points in a sheaf framework}\\

\vspace{1cm}

{\bf\Large Antonina N. Fedorova, Michael G. Zeitlin}\\

\vspace{1cm}

{\bf\large\it
IPME RAS, St.~Petersburg,
V.O. Bolshoj pr., 61, 199178, Russia}\\
\vspace{0.2cm}
{\bf\large\it e-mail: zeitlin@math.ipme.ru}\\
\vspace{0.2cm}
{\bf\large\it e-mail: anton@math.ipme.ru}\\
\vspace{0.2cm}
{\bf\large\it http://www.ipme.ru/zeitlin.html}\\
\vspace{0.2cm}
{\bf\large\it http://mp.ipme.ru/zeitlin.html}

\end{center}

\vspace{1cm}

\begin{abstract}
We consider some generalization of the theory of quantum states, which  
is based on the analysis of long standing problems and 
unsatisfactory situation with the possible interpretations 
of quantum mechanics. We demonstrate that the consideration of quantum states as sheaves
can provide, in principle, more deep understanding of some phenomena. 
The key ingredients of the proposed construction 
are the families of sections of sheaves with values 
in the category of the functional realizations of infinite-dimensional Hilbert 
spaces with special (multiscale) filtration. Three different symmetries, kinematical (on space-time), 
hidden/dynamical (on sections of sheaves), unified (on filtration of the full scale of spaces) 
are generic objects generating the full zoo of quantum phenomena (e.g., faster than light propagation). 
\end{abstract}

\vspace{15mm}

\begin{center}
{\large Submitted to Proc. of SPIE Meeting,}\\
\vspace{0.2cm}
{\large The Nature of Light: What are Photons? IV}\\
\vspace{0.2cm}
{\large San Diego, CA, August, 2011}

\vspace{5mm}

\end{center}
\end{tabular}
\end{center}
\newpage


\maketitle

\begin{abstract}
We consider some generalization of the theory of quantum states, which  
is based on the analysis of long standing problems and 
unsatisfactory situation with the possible interpretations 
of quantum mechanics. We demonstrate that the consideration of quantum states as sheaves
can provide, in principle, more deep understanding of some phenomena. 
The key ingredients of the proposed construction 
are the families of sections of sheaves with values 
in the category of the functional realizations of infinite-dimensional Hilbert 
spaces with special (multiscale) filtration. Three different symmetries, kinematical (on space-time), 
hidden/dynamical (on sections of sheaves), unified (on filtration of the full scale of spaces) 
are generic objects generating the full zoo of quantum phenomena (e.g., faster than light propagation).
\end{abstract} 
  
\keywords{Localization; quantum states; multiscales; hidden symmetry; sheaves.}


\section{Introduction}

During a relative long period, it is well-known 
that there is a great difference between (at least) the mathematical 
levels of the investigation of quantum phenomena in different regions.  
Really, in the area of the so-called Strings (Math)Physics, one needs all existing machinery, like 
deformation quantization,
non-commutative geometry, etc. [1] while more applicable 
in real life Quantum Mechanics/Physics uses old routines mainly.
Of course, Physics at Planckian scales demands a new vision regarding Physics of (realizable) 
Quantum Devices (like future CPU) 
but, at the same time, we deal with the same phenomenon (at least on qualitative level).
It is impossible to imagine that some hypothetical object
 can be more quantum or less quantum. It seems that 
it can be only quantum or not quantum, i.e. classical (surely, we ignore, at the moment, quasi-classics).
At the same time, even advanced modern Mathematics cannot help us in the
final (at least practically accepted) analysis of the long standing quantum phenomena and
the final classification of a zoo of interpretations. 
The well-known incomplete list is as follows [2]:

{\bf (L)\ }  entanglement, measurement, wave function collapse, 
decoherence, Copenhagen interpretation, consistent histories, 
many-worlds interpretation/multiverse (MWI), 
Bohm interpretation, ensemble interpretation, (Dirac) self-interference, ``instantaneous'' quantum interaction,
hidden variables, etc.

As a result, beyond  a lot of fundamental advanced problems at Planckian scales
we are still even unready to create 
the proper theoretical background for the reliable 
modeling and constructing of quantum devices far away from Planckian scales.

So, let us expose one more approach in an attempt to understand, 
in unified framework, some set of well-known phenomena.

\section{Quantum States/Patterns: Functions vs. Sheaves. Mathematical Side}

 Let us presume that Quantum Dynamics 
can be properly described as a part of Local Quantum Physics [3] 
via Wigner-Weyl-Moyal and/or Deformation Quantization approach which covers other ones. 
It means that:

{\bf i)} we use the language, machinery and ideology of the theory of pseudo-differential operators;

{\bf ii)} we work with the symbols of operators instead of the operators;

{\bf iii)} Quantum Evolution is described by (pseudo-differential) 
Wigner-like equations (Wigner-von Neumann-Moyal-Lindblad);

{\bf iv)} the adequate analysis of the full set of possible phenomena described by iii) 
demands the using of Microlocal Analysis
(although there are some phenomena which can be described more traditionally) [4]; 

{\bf v)} we need to consider Quantum State not as a function but as a sheaf [4]. 
It is more proper
(at least from formal, mathematical, point of view)
if we really want to take into account a lot of internal arguments from 
points i)-iv) above.

\section{Quantum States/Patterns: Functions vs. Sheaves. Physical Side, Hypothesis H1}

 It is very hard to believe that trivial simple solutions, like gaussians, can 
exhaust all variety of possible quantum states needed for the resolution of all contradictions, 
hidden inside the list {\bf (L)} mentioned above.
So, let us propose the following natural (physical) hypothesis:

\vspace{3mm}

{\bf (H1)\ } {\bf the physically reasonable really existing Quantum States cannot be described by means of functions.
Quantum state is a complex pattern which demands a set/class of functions/patches instead of one function 
for proper description and understanding.}

\vspace{3mm}

There is nothing unusual in {\bf (H1)\ } for physicists since Dirac's description of monopole.
All the more, there is nothing unusual for mathematicians who successfully used sheaves, germs, etc
in different areas.  
Definitely, the introducing of {\bf (H1)} causes a number of standard topics, the most important
of them  are motivations, formal (exact) definition and (at least) particular realizations.
Really, why need we to change our ideology after a century (since Planck) of success?
The answer is trivial and related to  the list {\bf (L)} which 
is overcompleted with contradictions and misunderstanding after many decades of discussions.

\section{On the Route to Right Description: States as Sheaves, Hypothesis H2}

In the following description, it is possible to find  some features or reminiscences 
of previous models and interpretations from the list {\bf L}, where
the most important points for allusions are the hidden parameters, 
localization, ensemble/statistical interpretation, MWI, Dirac's ``self-interference''.
Let us sketch the main ingredients.

{\bf 1)}.\ Arena for Quantum Evolution

First of all, we need to divide the kinematical and dynamical features of a set of 
Quantum States ($QS$).  
From the formal point of view it means that one needs to consider some bundle $(X, H, H_x)$
whose sections are the so-called $|\psi>$ 
functions or $QS$. Here $X$ is (kinematical) space-time 
base space with the proper 
kinematical symmetry group (like Galilei or Poincare ones), $H$ is a total formal Hilbert space 
and $H(x)=H_x$ are fibers with their own internal structures and hidden symmetries. 
In addition, such a bundle has the corresponding structure
group which connects different fibers. Of course, in a very particular case 
we have the constant bundle with the trivial structure group but non-trivial fiber symmetry. 
Anyway, as we shall demonstrate later, 
it is very reasonable 
to provide the one-to-one correspondence between Quantum States and 
the proper sections

\begin{eqnarray}
\qquad |QS> \ : \quad X \longrightarrow H,
 \qquad QS:\quad x \longmapsto H_x=H(x).
\end{eqnarray}

As a result, we have, at least,  three different symmetries inside this construction:

{\bf kinematical one on space-time, hidden one inside each fiber and the gauge-like 
structure group of the bundle as a whole.}

It is obvious that the kinematical laws (like relativity principles) depend on the 
proper type of symmetry and are absolutely different in the base space and in 
the fibers. 
It should be noted that the functional realizations of fibers and the total space 
are very important for our aims. Roughly speaking, it can be supposed that physical effects 
depend on the type of the particular functional realization of formal (infinite dimensional) 
Hilbert space. E.g., it is impossible to use infinite smooth approximations, like gaussians, 
for the reliable modeling of chaotic/fractal phenomena. So, the part of Physics at quantum scales 
is encoded in the details of the proper functional realization.

{\bf 2).\  Localization and a Tower of Scales}

It is well-known that  nobody can prove that 
gaussians (or even standard coherent states, etc) are an adequate 
and proper image for Quantum States really existing in the Nature.
We can suggest that at quantum scales other classes of functions or, 
more generally, other functional spaces (not $C^\infty$, e.g.) with the proper bases describe
the underlying physical processes.
There are two key features we are interested in. 

First of all, we need the best possible localization
properties for our trial base functions. 

Second, we need to take into account, in appropriate form, all contributions
from all internal hidden scales, from coarse-grained to finest ones.  
Of course, it is a hypothesis but it looks very reasonable:

{\bf (H2)\ } {\bf there is a (infinite) tower of internal scales in quantum region that 
             contributes to the really existing Quantum States and their evolution.}

\vspace{3mm}

So, we may suppose that the fundamental generating physical ``eigen-modes'' correspond 
to a selected functional realization 
and are localized in the best way. 
Let us note  the role of the proper hidden symmetries which are responsible for
the quantum self-organization and resulting complexity.

{\bf 3).\  An Ensemble of Scales: Self-interaction}

As a result of the description above, we may have non-trivial ``interaction'' 
inside an infinite hierarchy of modes or scales. 
It resembles, in some sense, 
a sort of turbulence or intermittency. Of course, here the generating avatar is a representation theory
of hidden symmetries which create the non-trivial dynamics of this ensemble of hierarchies.

{\bf 4).\  Hidden Parameters and Hidden Symmetry}

It is well-known that symmetries generate all things (at least) in fundamental physics.
Here, we have a particular case where the generic symmetry corresponds 
to the internal hidden symmetry of 
the underlying
functional realization. Moreover, as it is proposed above, we have even the more complicated structure 
because we believe
that $QS$ is not a function but a sheaf. As a result, we have interaction between two different 
symmetries, namely hidden symmetry in the fiber, that corresponds 
to the internal symmetry of the functional realization, 
and the structure
``gauge'' group of a sheaf, which provides multifibers transition/dynamics.
Both these algebraic structures can be parametrized by the proper group parameters which can play the role of 
famous ``hidden variables'' introduced many decades ago.

{\bf 5).\ MWI}

Of course, MWI or Multiverse interpretation can be covered by the structure sketched above.
Quantum States are the sections of our fundamental sheaf, so we can consider them as a 
collection of maps
between the patches of base space and fibers. All such maps simultaneously exist 
and, as an equivalence class, represent
the same Quantum State. We postpone the detailed description to the next Section but here let us mention that
each member of the full family can be considered as an object belonged to some fixed World.
Obviously, before measurement we cannot distinguish samples but after measurement we shall 
have the only copy in our hands.

\section{On the Way to Definition}

The main reason to introduce sheaves as a useful instrument for the analysis 
of Quantum States is related to
their main property which allows 
to assign to every region $U$ in space-time  
$X$ some family $F(U)$ of algebraic or geometric objects such as functions or differential operators.

The family can be restricted to smaller regions, and the compatible collections 
of families can be glued to give a family over larger regions, 
so it provides  connection between small and large scales, local and global data.

Informal construction is as follows.

Let $X$ be the space-time base space (some topological space)
with a system of open subsets $U \subset X$, then for every $U$ and map $F$ the image $F(U)$ is some 
object with internal structure (more generally, 
$F(U)$ takes values in some category ${\bf H}$) such that
for every two open subsets, $U$ and $V$, $V \subset U$ there is the so-called 
restriction map (more generally, morphism in the category ${\bf H}$), 
$r_V,_U: F(U) \to F(V)$ (restriction morphism).

A map $F$ will be a {\bf presheaf} if restriction morphism satisfies the following properties:

\vspace{5mm}

{\bf (a)} for every open subset $U \subset X$, the restriction morphism $r_{U,U} : F(U) \to F(U)$ 
is the identity morphism,

\vspace{5mm}

{\bf (b)} if there are three open subsets $W\subset V\subset U$, then $r_{W,V} r_{V,U} = r_{W,U}$.

\vspace{5mm}

This property 
provides the connection or ordering of the underlying scales.

In other words, 
let $O(X)$ be the category of open sets on $X$, 
whose objects are the open sets of $X$ and whose morphisms are inclusions. 

\vspace{3mm}

Then a presheaf ${\bf F}$ on $X$ with values in category ${\bf H}$ is the 
contravariant functor from $O(X)$ to ${\bf H}$.

\vspace{3mm}

$F(U)$ is called the section of ${\bf F}$ over $U$ and we consider it as  some pre-image for
adequate Quantum State $|QS>$.

\vspace{3mm}

But our goal, in this direction, is a {\bf sheaf}, so we need to add two additional properties.
Let $\{U_i\}_{i\in I}$ be some family of open subsets of $X$, $U = \cup_{i \in I} U_i$.

{\bf (c)} If $\Psi_1$ and $\Psi_2$ are two elements of $F(U)$ and $r_{U_i,U}(\Psi_1)=r_{U_i,U}(\Psi_2)$
for every $U_i$, then $\Psi_1=\Psi_2$.

\vspace{3mm}

{\bf (d)} for every $i$ let a section $\Psi_i \in F(U_i)$.  $\{\Psi_i\}_{i \in I}$ are compatible if, 
for all $i$ and $j$, 
$r_{U_i \cap U_j, U_i}(\Psi_i) = r_{U_i \cap U_j, U_j}(\Psi_j)$. 

\vspace{3mm}

For every set $\{\Psi_i\}_{i \in I}$ of compatible sections on $\{U_i\}_{i \in I}$, 
there exists the unique section $\Psi \in F(U)$ such that $r_{U_i,U}(\Psi) = \Psi_i$ 
for every ${i \in I}$.

\vspace{3mm}

The section ${\bf \Psi}$ is called the gluing of the sections ${\Psi_i}$.

Definitely, we can consider this property as allusion to the hypothesis 
of wave function collapse. 

Really, ${\bf \Psi}$ looks as Multiverse Quantum State Ensemble $\{\Psi_i\}$ while $\Psi_i$
is the result of
measurement in the patch $U_i$. And it is unique!

The next step is to specify the Quantum Category ${\bf H}$. According to our Hypothesis ${\bf H2}$,
we consider 
the category of the functional realization of (infinite-dimensional) Hilbert spaces with
proper filtration, which allows to take into account multiscale decomposition for all dynamical quantities
needed for the description of Quantum Evolution.

The well-known type of such filtration is the 
so-called multiresolution decomposition. 
It should be noted that the whole description is much more complicated because it demands the
consideration of both structures together, 
namely, the fiber structure generated by internal hidden symmetry 
of the chosen functional realization 
and the family of gluing sections ${\bf \Psi}$ in the unified framework.

\section
{Realization via Multiresolution: Dynamics, Measurement, Decoherence, etc.}

In the companion paper [7], we shall consider in details 
one important realization of this construction based on the 
local nonlinear harmonic analysis which has, as the key ingredient, the so-called 
Multiresolution Analysis (MRA) [5]. 
It allows us to describe internal hidden dynamics on a tower of scales [5], [6].
Introducing the Fock-like space structure on the whole space of internal hidden scales,
we have the following MRA decomposition (refs. [9]-[22] for related methods, approaches, models):

\begin{eqnarray}
H=\bigoplus_i\bigotimes_n H^n_i
\end{eqnarray}
for the set of n-partial Wigner functions (states):

\begin{equation}
W^i=\{W^i_0,W^i_1(x_1;t),\dots,
W^i_N(x_1,\dots,x_N;t),\dots\},
\end{equation}
where
$W_p(x_1,\dots, x_p;t)\in H^p$,
$H^0=C,\quad H^p=L^2(R^{6p})$ (or any different proper functional spa\-ce), 
with the natural Fock space like norm: 
\begin{eqnarray}
(W,W)=W^2_0+
\sum_{i}\int W^2_i(x_1,\dots,x_i;t)\prod^i_{\ell=1}\mu_\ell.
\end{eqnarray}

Now we consider some phenomenological description which presents 
some attempt of the qualitative description of quantum dynamics 
as a whole and in comparison with its classical counterpart.
It is possible to take, for reminiscence, 
the famous Dirac's phrase that ``an electron can interact only 
itself via the process of quantum interference''.

Let $G$ be a hidden/internal symmetry group
on the spaces of Quantum States, which generates, via MRA,
the multiscale/multiresolution representation for all 
dynamical quantities, unified in object $O(t)$, such as states, observables, 
partitions (e.g., Wigner quasi-distributions): 
$O^i(t)=\{\psi^i(t), Op^i(t), W_n^i(t)\}$, where $i$ is the proper scale index.

Then, the following commutative diagram represents the details of quantum life from 
the point of view of the representations of $G$ on the chosen functional realization which leads to
the decomposition of the whole quantum evolution into the orbits or scales corresponding 
to the proper level of resolution. Morphisms $W(t)$ describe Wigner-Weyl evolution in the 
algebra of symbols, while the processes of interactions with open World, such as
the measurement or decoherence, correspond to morphisms (or even functors) $m(t)$
which transform the infinite set of scales,
characterizing the quantum object, into finite ones, sometimes consisting of one element
(demolition/destructive measurement)

\begin{eqnarray}
&&\qquad\qquad\qquad W(t) \nonumber\\
&&\lbrace O^i(t_1) \rbrace \qquad  \longrightarrow   \qquad \lbrace O^j(t_2) \rbrace \nonumber\\
&&\nonumber\\
&&\downarrow m(t_1) \qquad \ \qquad \qquad  \downarrow m(t_2) \nonumber\\
&&\qquad\qquad\qquad\widetilde{W(t)}\nonumber\\
&&\lbrace O^{i_c}(t_1) \rbrace \qquad  \longrightarrow  \qquad \lbrace O^{j_c}(t_2) \rbrace,\nonumber
\end{eqnarray}

\vspace{3mm}

where the reduced morphisms $\widetilde{W(t)}$ correspond to (semi)classical or 
quasiclassical evolution.


Qualitatively, 

{\bf Quantum Objects}

can be represented by an infinite or sufficiently large set of coexisting and
interacting subsets 

\vspace{3mm}

while 

\vspace{3mm}

{\bf (Quasi)Classical Objects} can be described by one or a few only levels of resolution
with (almost) suppressed interscale self-interaction.

It is possible to consider Wigner functions as some measure of the quantum character of the system:
as soon as it becomes positive, we arrive to classical regime and so there is 
no need to consider the full hierarchy decomposition in the MRA representation.

So, Dirac's self-interference is nothing else than the multiscale mixture/intermittency.

Certainly, the degree of this self-interaction leads to different qualitative types 
of behaviour, such as localized quasiclassical states, separable, entangled, chaotic etc.
At the same time, the instantaneous quantum interaction or transmission of 
(quantum) information from Alice 
to Bob takes place not in the physical kinematical space-time but in Hilbert spaces of 
Quantum States 
in their proper functional realization where there is a different kinematic life.

To describe a set of Quantum Objects, we need to realize our Space of States (Hilbert space)
not as one functional space but as the so-called and well known in 
mathematics, scale of spaces, e.g. $B^s_{p,q}$, $F^s_{p,q}$.
The proper multiscale decomposition for the scale of spaces provides us by the 
method of the description of a set of quantum objects in case if the ``size'' of 
one Hilbert space of states is not enough to describe the complicated internal World.
We will consider it elsewhere, while here we consider the one-scale case
(to avoid possible misunderstanding, we need to mention that one-scale case 
is also described by an infinite scale of spaces, but it is the internal 
decomposition of the unique, attached to the problem, Hilbert space).
Definitely, the full family of sections of non-trivial sheaf, as a model of $QS$,  demands  
to take into account the double-hierarchy of the underlying internal scales generated by means of
the corresponding hidden symmetries. 
As a result, on the proper orbits, we have nontrivial entangled dynamics, 
especially in contrast with its ``classical'' quantum orthodox counterpart.


\section{Discussions and Summary}
\begin{center}

\vspace{3mm}

{\Large\bf How to get rid of geometrical points from Quantum Physics}\\
\end{center}

(Reasonable) Questions:

\noindent{\bf Question 1.} 

Is a geometrical point a good image/model for Quantum Object?

If the answer on the Q1 is "possible, not'' or "definitely, not"
then the next Question, Q2, is a direct consequence of Q1.

\noindent{\bf Question 2.} 

Is a (point) function (standard, usual function) a proper model for $\psi$
function and related objects?

If the answer is not, then we have Hypothesis 1:

\noindent{\bf Hypothesis 1.}

Quantum Objects need to be defined on open sets (in standard topological
sense) or some system of open sets, or, to be more
precise mathematically, filtrations instead of one-point-sets
(geometrical points)

\vspace{3mm}
   
Then, Hypothesis 2 will be obvious:

\noindent{\bf Hypothesis 2.}

Sheaves are the proper realization for modeling Quantum States/Objects


One step more:

\noindent{\bf Question 3.}

Is the kinematics of space-time manifolds with a proper symmetry group, like
Poincare, e.g., a generic object for understanding
Quantum Paradoxes/Interpretations, such as instantaneous interactions,
nonlocality, hidden parameters, etc.?

Moreover, can (special) Relativity Theory describe, in principle,
instantaneous interaction between Quantum Objects?

\vspace{3mm}

If the answer is not, then

\noindent{\bf Hypothesis 3.}

The most nontrivial Quantum Phenomena can be described by means of
underlying hidden symmetry of functional realization of Hilbert
Space of States generating the full Zoo of Quantum Phenomena.

Of course, there is no contradiction with trivial kinematical laws
(relativity): these are two different coexisting objects.

 \vspace{3mm}  

{\bf The Full Generic Symmetry is a Unification of these Symmetries.}

\vspace{3mm}

E.g., the instantaneous quantum interaction is a phenomenon of Function
Space Kinematics (with proper hidden symmetry) but not an
attribute of Minkowskian kinematics. 

 \vspace{3mm}  

Roughly (physically) speaking, the Quantum Signal between Quantum Objects (in the hands
of Alice and Bob) propagates along Functional Realization of Hilbert space
(with its own kinematis, symmetry, etc.) but not along Minkowskian
space-time with Einsteinian Relativity and underlying Poincare symmetry.

\vspace{3mm}

Roughly (mathematically) speaking:
Physical Laws at Base Space (Space--Time Manifold) of Generic Fiber Bundle
(or Sheaf) are different from Physical Laws at fibers/slices/sections and,
moreover, from Laws proper for the Total Space.

\vspace{3mm}
For the final analysis we need to unify all underlying (mostly hidden) generic
Symmetries to have a chance for  right description.

\vspace{3mm} 

That is our point of view.

\vspace{3mm}

(Roughly) mathematically speaking, we can summarize our approach in a form of diagram:

\section{Paradigm for a Quantum State Representation}

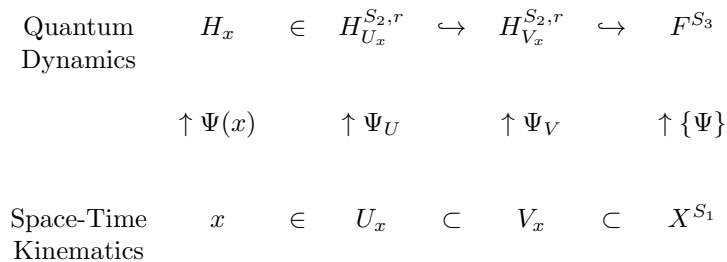
\vspace{5mm}

\begin{center}
\begin{tabular}{cccccccc}
{\rm Quantum} & $H_x$     &  $\in$    &$H^{S_2,r}_{U_x}$ &$\hookrightarrow$     &$H^{S_2,r}_{V_x}$  &$\hookrightarrow$ &$F^{S_3}$\\
{\rm Dynamics}&&&&&&&\\
&&&&&&&\\
&$\uparrow\Psi(x)$ &      &$\uparrow\Psi_U$&          &$\uparrow\Psi_V$ &          &$\uparrow\{\Psi\}$\\
&&&&&&&\\
&&&&&&&\\
{\rm Space-Time} &$x$               &$\in$ &$U_x$           &$\subset$ &$V_x$            &$\subset$ &$X^{S_1}$\\
{\rm Kinematics}&&&&&&&\\
&&&&&&&\\
\end{tabular}
\end{center}

\vspace{3mm}

Here $S_1, S_2, S_3$ are the Three  Different Generic Symmetries generating Three Different Sets of Physical Laws 
on the Base (Kinematical) Space $X$, on the Fibers (Local Hilbert Spaces over local domains of Base Space),  $H^{S_2,r}_{U_x}$, 
and on the Total Quantum Hilbert Space/Quantum Sheaf, $F^{S_3}$. 
It should be remembered that Laws of Kinematics (Einstein or Galileo Relativity), are attributes of the Base Space $X$ only while  
Quantum Laws need to be described by Total Quantum Hilbert Space and the full diagram above.
So, the standard ``local''/ ``point function'' $\Psi$ Function (left row) is not enough to restore properly 
a full set of quantum data corresponding to Quantum State (QS).
Instead of that, we consider the full family of sections over open domains unified in a whole Quantum Sheaf 
as a correct representative of the real QS. Correspondingly, the resulting ``microlocal/sheaf description'' [4] has
a lot of additional possibilities to store full quantum information. As a result, Quantum Dynamics in such a framework is more 
complicated and complex than in the standard ``orthodox'' case. So, in such a way we open new horizons and, at least, we
get rid of the structureless geometrical points from Quantum Context. Details will be considered elsewhere [23].

\section{Conclusions}

It seems very reasonable that there are no chances for the solution of long standing problems
and novel ones if we constraint ourselves by old routines and the old zoo of simple solutions like
gaussians, coherent states and all that.

Evidently, that even the mathematical background of regular Quantum Physics demands  
new interpretations and approaches. Let us mention only the procedures of quantization 
as a generic example.

In this respect, we can hope that our sheaf extension for representation of $QS$, 
which is natural from the formal point of view, may be very productive 
for the more deep understanding of the underlying (Quantum) Physics, especially, if we consider it 
together
with the category of multiscale filtered functional realizations decomposed into the entangled orbits 
generated by actions
of internal hidden symmetries. In such a way, we open a possibility 
for the exact description of a lot of phenomena 
like entanglement and measurement, wave function collapse, 
self-interference, instantaneous quantum interaction, Multiverse, 
hidden variables, etc. In the companion paper [7] we consider the machinery needed for 
the generation of a zoo of the complex quantum patterns during  Wigner-Weyl evolution. 
  

\acknowledgements

We are very grateful to Prof. Chandrasekhar Roychoudhuri and Michael 
Ambroselli (University of Connecticut, Storrs),
and Matthew Novak (SPIE) for their kind 
attention and help provided our presentations during 
SPIE2011 Meeting ``The Nature of Light: What are Photons? IV''
at San Diego. 
We are indebted to Dr. A. Sergeyev (IPME RAS/AOHGI) for his encouragement.


\begin{thebibliography}{23}


\bibitem{1}
A. Connes, M. Marcolli, {\it Noncommutative Geometry, Quantum Fields and Motives}, AMS, (2009).

\bibitem{2}
http://en.wikipedia.org/wiki/Interpretation of quantum mechanics and references therein.

\bibitem{3}
R. Haag, {\it Local Quantum Physics}, Springer, (1992).

\bibitem{4}
M. Kashiwara, P. Schapira, {\it Sheaves on Manifolds}, Springer, (1994).


\bibitem{5}
Y. Meyer, {\it Wavelets and Operators}, Cambridge Univ. Press, (1990).

\bibitem{6}
H. Triebel, {\it Theory of Functional Spaces}, Birkhauser, (1983).

\bibitem{7}
A.N. Fedorova and M.G. Zeitlin,
Quantum points/patterns, Part 2.
From quantum points to quantum patterns via multiresolution, this Volume.


\bibitem{8}
A.N. Fedorova and M.G. Zeitlin,
Quasiclassical Calculations for Wigner Functions via Multiresolution,
Localized Coherent Structures and Patterns Formation in Collective Models of Beam Motion, in
{\it Quantum Aspects of Beam Physics}, Ed. P. Chen,
World Scientific, Singapore, pp.~527--538, 539--550 (2002);
arXiv: phy\-sics\-/0101006; physics/0101007.

\bibitem{9}
A.N. Fedorova and M.G. Zeitlin,
BBGKY Dynamics: from Localization to Pattern Formation,
in {\it Progress in Nonequilibrium Green's Functions II}, Ed. M. Bonitz,
World Scientific, pp.~481--492 (2003)
arXiv: physics/0212066.

\bibitem{10}
A.N. Fedorova and M.G. Zeitlin,
Pattern Formation in Wigner-like Equations via Multiresolution, in
{\it Quantum Aspects of Beam Physics}, Eds. Pisin Chen, K. Reil, World
Scientific,  pp.~22-35 (2003); Preprint SLAC-R-630;
arXiv: quant-phys/0306197.

\bibitem{11}
A.N. Fedorova and M.G. Zeitlin,
Localization and pattern formation in Wigner representation via multiresolution,
{\it Nuclear Inst. and Methods in Physics Research, A}, {\bf 502A/2-3}, pp.~657 - 659 (2003);
arXiv: quant-ph/0212166.

\bibitem{12}
A.N. Fedorova and M.G. Zeitlin,
Fast Calculations in Nonlinear Collective Models of Beam/Plasma Physics,
{\it Nuclear Inst. and Methods in Physics Research, A}, {\bf 502/2-3}, pp.~660 - 662 (2003);
arXiv: physics/0212115.



\bibitem{13}
A.N. Fedorova and M.G. Zeitlin,
Classical and quantum ensembles via multiresolution: I-BBGKY hierarchy;
Classical and quantum ensembles via multiresolution. II. Wigner ensembles;
{\it Nucl. Instr. Methods Physics Res.}, {\bf 534A}, pp.~309-313; 314-318 (2004);
arXiv: quant-ph/0406009; quant-ph/0406010.


\bibitem{14}
A.N. Fedorova and M.G. Zeitlin,
Localization and Pattern Formation in Quantum Physics. I. Phenomena of Localization,
in {\it The Nature of Light: What is a Photon?}
{\it SPIE}, vol.{\bf 5866}, pp.~245-256 (2005); arXiv: quant-ph/0505114;


\bibitem{15}
A.N. Fedorova and M.G. Zeitlin,
Localization and Pattern Formation in Quantum Physics. II. Waveletons in Quantum Ensembles,
in {\it The Nature of Light: What is a Photon?}{\it SPIE}, vol. {\bf 5866}, pp.~257-268 (2005);
arXiv: quant-ph/0505115.


\bibitem{16}
A.N. Fedorova and M.G. Zeitlin,
Pattern Formation in Quantum Ensembles,
{\it Intl. J. Mod. Physics} {\bf B20}, pp.~1570-1592 (2006);
arXiv: 0711.0724.

\bibitem{17}
A.N. Fedorova and M.G. Zeitlin,
Patterns in Wigner-Weyl approach, Fusion modeling in plasma physics: Vlasov-like systems,
{\it Proceedings in Applied Mathematics and Mechanics (PAMM)}, Volume {\bf 6},
Issue 1, pp.~625-626, pp.~627-628, Wiley InterScience, (2006).


\bibitem{18}
A.N. Fedorova and M.G. Zeitlin,
Localization and Fusion Modeling in Plasma Physics. Part I: Math Framework
for Non-Equilibrium Hierarchies,
pp.61-86, in {\it Current Trends in International Fusion Research}, Ed. E. Panarella, R. Raman,
National Research Council (NRC) Research Press, Ottawa, Ontario, Canada (2009);
arXiv: physics/0603167.

\bibitem{19}
A.N. Fedorova and M.G. Zeitlin,
Localization and Fusion Modeling in Plasma Physics. Part II: Vlasov-like Systems. Important Reductions,
pp.87-100, in {\it Current Trends in International Fusion Research}, Ed. E. Panarella, R. Raman,
National Research Council (NRC) Research Press, Ottawa, Ontario, Canada (2009);
arXiv: physics/0603169.

\bibitem{20}
A.N. Fedorova and M.G. Zeitlin,
Fusion State in Plasma as a Waveleton (Localized (Meta)-Stable Pattern),
p.~272, in
{\it AIP Conference Proceedings}, Volume {\bf 1154}, Issue 1,
{\it Current Trends in International Fusion Research}, Ed. E. Panarella, R. Raman, AIP (2009).

\bibitem{21}
A.N. Fedorova and M.G. Zeitlin,
Exact Multiscale Representations for (Non)-Equilibrium Dynamics of Plasma,
p.~291, in
{\it AIP Conference Proceedings}, Volume {\bf 1154}, Issue 1,
{\it Current Trends in International Fusion Research}, Ed. E. Panarella, R. Raman, AIP (2009).

\bibitem{22}
A.N. Fedorova and M.G. Zeitlin,
Fusion Modeling in Vlasov-Like Models,
{\it J.Plasma Fusion Res. Series}, Vol. {\bf8}, pp.~126-131 (2009).

\bibitem{23}
A.N. Fedorova and M.G. Zeitlin, Quantum Sheaves, in progess.

\end{thebibliography}
\end{document}